# Self-Adaptive Binary-Addition-Tree Algorithm-Based Novel Monte Carlo Simulation for Binary-State Network Reliability Approximation

Wei-Chang Yeh
Department of Industrial Engineering and Engineering Management
National Tsing Hua University
P.O. Box 24-60, Hsinchu, Taiwan 300, R.O.C.
yeh@ieee.org

*Abstract* — The Monte Carlo simulation (MCS) is a statistical methodology used in a large number of applications. It uses repeated random sampling to solve problems with a probability interpretation to obtain high-quality numerical results. The MCS is simple and easy to develop, implement, and apply. However, its computational cost and total runtime can be quite high as it requires many samples to obtain an accurate approximation with low variance. In this paper, a novel MCS, called the self-adaptive BAT-MCS, based on the binary-adaption-tree algorithm (BAT) and our proposed self-adaptive simulation-number algorithm is proposed to simply and effectively reduce the run time and variance of the MCS. The proposed self-adaptive BAT-MCS was applied to a simple benchmark problem to demonstrate its application in network reliability. The statistical characteristics, including the expectation, variance, and simulation number, and the time complexity of the proposed self-adaptive BAT-MCS are discussed. Furthermore, its performance is compared to that of the traditional MCS extensively on a large-scale problem.

# 1. INTRODUCTION

Monte Carlo methods (MCSs) are a broad class of computational algorithms and statistical methodologies that are too complex to solve analytically [1]. MCSs play a crucial role in enormous modern and dynamics complex applications including engineering [2], physical sciences [3], climate change and radiative forcing [4], computational biology [5], computer graphics, applied statistics [6], artificial intelligence for games [7], design and visuals [8], search and rescue [9], finance and business, and law [10].

Various binary-state networks are used in real-world applications such network resilience [11], wireless sensor network [12], pipe networks [13], the internet of things [14], and social networks [15], and such networks are constantly growing. For example, social networks have more than 27 billion users [16]. Network reliability, an important index for evaluating network performance, represents the probability that a network is currently functioning successfully. However, it is NP-hard to calculate the exact binary-state network reliability [17, 18, 19, 20], making the MCS an effective and popular tool for estimating the reliability of all types of networks.

The basic idea behind using the MCS to estimate binary-state network reliability is to generate enough numbers randomly for all network components to obtain simulated results for a certain number of replications [21]. The results are then averaged to obtain an approximate reliability value for decision support in modern networks, which change dynamically and rapidly [22, 23].

In general, the MCS is easier to use than an exact-solution algorithm for obtaining the network reliability [24]. However, the results obtained from the MCS are not real answers but approximate reliabilities with a variance, which decreases with increasing simulation number [25, 26]. Moreover, the MCS requires a random number for each component in a simulation, and multiple replications for each design alternative affects its efficiency [27, 28]. If the number of random variates is large, the total cost of MCS can be exorbitantly high [1, 8]. Therefore, it is necessary to develop more efficient and robust MCSs through optimization for efficiency while calculating an acceptable approximate reliability [26, 27].

The binary-addition-tree algorithm (BAT) is an emerging search method for solving optimization problems [28]. The BAT has been implemented to calculate the exact reliability of different types of networks directly [28, 29, 30] and solve various practical problems, including resilience assessment [31], predicting the spread of wildfires [32] and propagation probability of computer viruses [33].

Because BAT is easy to program, flexible to make-to-fit, and more efficient than traditional search methods, including the depth-first search (DFS) [34], breadth-first search (BFS) [35], and universal generating function methodology (UGFM) [36], a new MCS called the BAT-MCS is proposed based on the self-adaptive BAT by integrating BAT [28], MCS [25], and our proposed self-adaptive simulation-number algorithm to obtain a robust estimator of the binary-state network reliability efficiently.

Our study aimed to develop a simple, efficient, and robust MCS based on the self-adaptive BAT. We experimentally show that the proposed self-adaptive BAT-MCS can overcome the NP-hard obstacle by finding an accurate and robust solution efficiently rather than finding an exact reliability value.

The remainder of this paper is organized as follows. Section 2 introduces the required acronyms, notations, nomenclature, and assumptions used in this study. Section 3 introduces the two fundamental modules of the proposed self-adaptive BAT-MCs, BAT [28] and MCS [25], together with the path-based layered search algorithm (PLSA) [37]. Section 4 discusses the proposed novel self-adaptive simulation-number algorithm, lower and upper bounds of each super vector, pseudocode to combine the self-adaptive BAT and MCS, and statistical characteristics of the self-adaptive BAT-MCS. Section 5 presents the step-by-step implementation of the proposed self-adaptive BAT-MCS based on a simple benchmark network. Section 5 also describes the computational experiment performed on a large-scale network to analyze the relationships among the estimators, number of coordinates in the super vector obtained from the BAT, number of simulations in the MCS using the proposed self-adaptive simulation-number algorithm, and component reliabilities. Finally, Section 6 concludes the paper.

**2. ACRONYMS, NOTATIONS, NOMENCLATURE, AND ASSUMPTIONS**

Important acronyms, notations, assumptions, and nomenclatures related to the proposed BAT-MCS are presented in this section.

**2.1 Acronyms**

MCS : Monte Carlo simulation

BAT : Binary-Addition-Tree Algorithm

Self-Adaptive BAT-MCS   the proposed MCS based on BAT with self-adaptive simulation-number algorithm

LSA: Layered search algorithm

PLSA: Path-based LSA

## 2.2 Notations

$|\bullet|$: number of elements in $\bullet$

$\|\bullet\|$: number of coordinates in vector $\bullet$

$E[\bullet]$: expectation value of estimator $\bullet$

$Var[\bullet]$: variance of estimator $\bullet$

$Pr(\bullet)$: probability to have $\bullet$

$Pr^*(\bullet)$: approximate probability to have $\bullet$

$a_i$: arc $i$

$p$: $Pr(a) = p$ for all arc $a$

$V$: $V = \{1, 2, \ldots, n\}$

$E$: $E = \{a_1, a_2, \ldots, a_m\}$

$X$: (arc) state vector

$X(a_i)$: value of the $a_i$ in $X$ for $i = 1, 2, \ldots, m$, e.g., $X(a_1) = X(a_2) = X(a_3) = 1$ and $X(a_4) = X(a_5)$ =0 if $X = (1, 1, 1, 0, 0)$.

$\mathbf{D}$: arc state distribution listed $\mathbf{D}(a) = Pr(a)$ for all $a \in E$, e.g., $\mathbf{D}$ is listed in **Table 1**.

Table 1. Binary-state arc distribution **D**

| $i$ | $\mathbf{D}(a_i) = Pr(a_i)$ |
|---|---|
| 1 | 0.9 |
| 2 | 0.8 |
| 3 | 0.7 |
| 4 | 0.6 |
| 5 | 0.5 |

$G(V, E)$: an undirected graph with $V$ and $E$. For example, **Figure 1** is a graph with $V = \{1, 2, 3, 4\}$, $E = \{a_1, a_2, a_3, a_4, a_5\}$, source node 1, and sink node 4 $G(V, E, \mathbf{D})$:    $G(V, E)$ with the **D**. For example, **Figure 1** is a binary-state network if **D** is as given in **Table 1**.

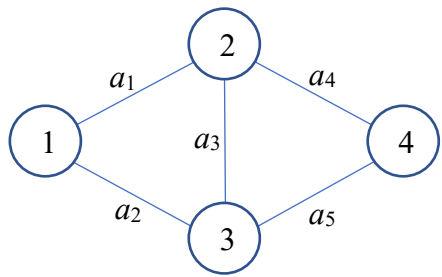

Figure 1. Example network

$G(X)$:    subgraph corresponding to $X$ such that $G(X) = G(V, \{a \in E \mid \text{for all } a \text{ with } X(a) = 1\}, \mathbf{D})$. For example, $G(X)$ is depicted in **Figure 2**, where $X = (1, 1, 1, 0, 0)$ in **Figure 1**.

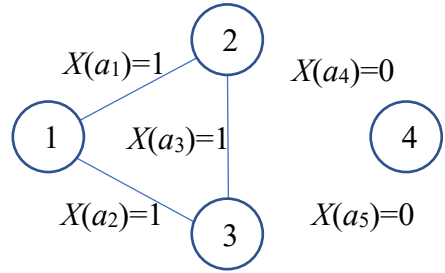

Figure 2. $X = (1, 1, 1, 0, 0)$ and $G(X)$, where $G(V, E)$ is shown in **Figure 1**.

$\mathbf{D}_S$:    state distribution based on vector $S$ such that $\mathbf{D}_S(a_j) = \begin{cases} \mathbf{D}(a_j) & \text{if } j > \|S\| \\ 1 & \text{if } j \leq \|S\| \text{ and } S(a_j) = 1 \\ 0 & \text{otherwsie} \end{cases}$

$G(\mathbf{D}_S)$:    subgraph corresponding to state distribution $\mathbf{D}_S$ such that $G(\mathbf{D}_S) = G(V, E, \mathbf{D}_S)$.

$R$:    exact reliability

$R_{\text{MCS}}$:    approximate reliability obtained from the MCS

$R_{\text{MCS},j}$:    approximate reliability obtained from the $j$th simulation of the MCS

$R_{\text{BAT-MCS}}$:    approximate reliability obtained from the proposed self-adaptive BAT-MCS

$R(G)$:    reliability of $G(V, E, \mathbf{D})$ and $R(G) = R$

$R^*(G(X))$:    approximate reliability of $G(X)$

$X \ll X^*$:    obtained sequence of vector $X$ is earlier than that of vector $X^*$ in BAT

$X \underset{\sim}{\ll} X^*$:    obtained sequence of vector $X$ is earlier than or equal to that of vector $X^*$ in BAT

$X < X^*$:    $X(a) < X^*(a)$ for all $a \in E$

$X \leq X^*$:    $X(a) \leq X^*(a)$ for all $a \in E$

$\delta$:    number of coordinates in super vectors obtained from the self-adaptive BAT-MCS

$\Omega$:    $\{X \mid X \text{ is a } m\text{-tuple vector}\}$

$\Omega_\delta$:    $\{S \mid S \text{ is a super vector with } \|S\| = \delta\}$

$\Omega(S)$:    $\{X \mid X \in \Omega \text{ with } X(a_i) = S(a_i) \text{ for all } i = 1, 2, \ldots, \|S\| \text{ and } S \in \Omega_\delta\}$

$L(S)$:    $L(S) = X \in \Omega$ such that $X(a_i) = S(a_i)$ for all $i = 1, 2, \ldots, \|S\|$ and $X(a_i) = 0$ for all $i = \|S\|+1, \|S\|+2, \ldots, m$, and $S \in \Omega_\delta$.

$U(S)$:    $U(S) = X \in \Omega$ such that $X(a_i) = S(a_i)$ for all $i = 1, 2, \ldots, \|S\|$, $X(a_i) = 1$ for all $i = \|S\|+1, \|S\|+2, \ldots, m$, and $S \in \Omega_\delta$.

$L_\delta$:    $\{S \mid S \in \Omega_\delta$ and $L(S)$ is connected$\} \subseteq \Omega_\delta$.

$U_\delta$:    $\{S \mid S \in \Omega_\delta$ and $U(S)$ is disconnected$\} \subseteq \Omega_\delta$.

$F_\delta$:    $F_\delta = \Omega_\delta - (L_\delta \cup U_\delta)$

$N_{sim}$:    number of simulations in each run of MCS or BAT-MCS

$N_{run}$:    number of runs for each MCS or BAT-MCS

$\varepsilon$    error between the estimator and the exact solution

$\sigma$    standard deviation

$T$    runtime

## 2.3 Nomenclature

Binary-state network:    A network with binary-state arcs only, that is, the state of each arc is either zero or one.

Reliability:    The probability that a network is operating successfully.

Super vector:    $S = (s_1, s_2, \ldots, s_\delta)$ is a sub-vector such that $S(a_i) = s_i$ and $i = 1, 2, \ldots, \delta$ and $\delta \leq m$. [30]

Connected vector:    $X$ is a connected vector if nodes 1 and $n$ are connected in $G(X)$.

Disconnected vector:    $X$ is a disconnected vector if nodes 1 and $n$ are disconnected in $G(X)$.

## 2.4 Assumptions [38]

1. Networks have no parallel arcs or loops.
2. Nodes are all completely reliable and connected.
3. Arcs have two states and all states are statistically independent.

## 3. REVIEW OF BAT, PLSA, AND MCS

The proposed self-adaptive BAT-MCS uses the BAT [28] to generate super vectors [30], PLSA to filter out all feasible super vectors by verifying their connectivity [37], and MCS to calculate the approximate probability of each feasible super vector [24, 25]. Hence, the BAT, PLSA, and MCS and their important statistical characteristics are reviewed briefly in this section.

**3.1 BAT**

A binary-state vector has $m$ binary-state coordinates, and the value of each coordinate is either zero or one, denoting that the corresponding arc fails or works, respectively. The (binary) BAT proposed by Yeh [28] aims to generate all possible $m$-tuple binary-state vectors, regardless of whether it is feasible or infeasible based simply on the binary addition operator.

The procedure adopted by the BAT to generate all vectors is analogous to adding one to a binary code from the first coordinate and gradually moving to the last coordinate [28, 30] in the binary-state vector. The general procedure for BAT is explained in the following pseudocode [28]:

**Algorithm: BAT**

**Input:**   $m$.

**Output:**   All $2^{|m|}$ $m$-tuple binary-state vectors.

**STEP B0.** Let $i = 1$ and $X$ be a $m$-tuple vector zero.

**STEP B1.** If $X(a_i) = 0$, let $X(a_i) = 1$, $i = 1$, a new $X$ is generated, and return to STEP B1.

**STEP B2.** If $i < m$, let $X(a_i) = 0$, $i = i + 1$, and return to STEP B1.

Let $i$ be the index of the current coordinate and $X$ be the current vector. STEP B0 initializes $i$ to one and all values of the coordinates in the current vector $X$ to zero. The main body of the BAT is the loop from STEP B1 to STEP B2, and it updates $X$ repeatedly to generate each vector by adding one to $X(a_i)$, that is, the current coordinate.

The value of $X(a_i)$ is changed to one and a new $X$ is found in the loop if $X(a_i) = 0$. Otherwise, $X(a_i)$ is changed from one to zero and moved to the next coordinate to repeat the above process until $X(a_i)$ is changed from zero to one. If $i = m$, this loop terminates because the current $X$ becomes a zero vector, and all the vectors are already found.

The BAT is very simple because its pseudocode has only three statements. The time complexity of the BAT has been proven to be $O(2^{m+1})$ in [39]. Hence, as $O(2^{m+1})$ and the total number of vectors is $2^m$, the BAT is very efficient in generating all vectors. In addition, vector $X$ is updated repeatedly without introducing any new vector, that is, the BAT is space-efficient [28]. In addition, the function $F(X)$ can be made-to-fit, for example, the cost, probability [32], time [40], weight, etc., based on the problem and calculate $F(X)$ after a new $X$ obtained in STEP B1, that is, the BAT is easy to customize [30].

Thus, the BAT is easy to code, flexible to make-to-fit, and efficient in both runtime and space. Experiments [30] showed that the BAT outperforms other well-known search methods, such as the DFS [34], BFS [35], and UGFM [36]. Hence, many versions and of the BAT have emerged applications [28, 29, 30, 31, 32, 33, 39, 40].

For example, in the five-arc binary-state network shown in the example network in **Figure 1,** , each vector is a 5-tuple, the first vector is $X = (0, 0, 0, 0, 0)$, and $i = 1$ in STEP B0. Using binary addition, by adding one to the current coordinate, that is, the first coordinate, the new $X = (1, 0, 0, 0, 0)$ is obtained as the second vector and the new $i$ is still equal to zero in STEP B1. After the new $X$ is generated, there is no need to store the current $X$, thereby saving computer memory.

If the cost functions $C(\bullet)$, time function $T(\bullet)$, weight function $W(\bullet)$, and probability $\Pr(\bullet)$ of each arc are 8, 7, 6, and 0.9, respectively, then $C(X) = 8$, $T(X) = 7$, $W(X) = 6$, and $\Pr_{0.9}(X) = 0.9 \times 0.1^4 = 0.00009$, for the new $X = (1, 0, 0, 0, 0)$. Note that the function we need to define and calculate depends on the problem.

Because $i = 1$ and $X(a_1) = 1$ in $(1, 0, 0, 0, 0)$, we need to change $X(a_1)$ to zero, move to the next coordinate by changing $i$ from 1 to 2, and return to STEP B1 to repeat the same process until the current coordinate value is changed to 1 to generate a new $X$. The complete process of the BAT illustrated in **Figure 1** is presented in **Table 2**.

**Table 2.** All vectors obtained from the BAT.

| iteration | $X$ | $C(X)$ | $T(X)$ | $W(X)$ | $\Pr_{0.9}(X_i)$ |
|---|---|---|---|---|---|
| 1 | (0, 0, 0, 0, 0) | 40 | 35 | 30 | 0.59049 |
| 2 | (1, 0, 0, 0, 0) | 32 | 28 | 24 | 0.06561 |
| 3 | (0, 1, 0, 0, 0) | 32 | 28 | 24 | 0.06561 |
| 4 | (1, 1, 0, 0, 0) | 24 | 21 | 18 | 0.00729 |
| 5 | (0, 0, 1, 0, 0) | 32 | 28 | 24 | 0.06561 |
| 6 | (1, 0, 1, 0, 0) | 24 | 21 | 18 | 0.00729 |
| 7 | (0, 1, 1, 0, 0) | 24 | 21 | 18 | 0.00729 |
| 8 | (1, 1, 1, 0, 0) | 16 | 14 | 12 | 0.00081 |
| 9 | (0, 0, 0, 1, 0) | 32 | 28 | 24 | 0.06561 |
| 10 | (1, 0, 0, 1, 0) | 24 | 21 | 18 | 0.00729 |
| 11 | (0, 1, 0, 1, 0) | 24 | 21 | 18 | 0.00729 |
| 12 | (1, 1, 0, 1, 0) | 16 | 14 | 12 | 0.00081 |
| 13 | (0, 0, 1, 1, 0) | 24 | 21 | 18 | 0.00729 |
| 14 | (1, 0, 1, 1, 0) | 16 | 14 | 12 | 0.00081 |
| 15 | (0, 1, 1, 1, 0) | 16 | 14 | 12 | 0.00081 |
| 16 | (1, 1, 1, 1, 0) | 8 | 7 | 6 | 0.00009 |
| 17 | (0, 0, 0, 0, 1) | 32 | 28 | 24 | 0.06561 |
| 18 | (1, 0, 0, 0, 1) | 24 | 21 | 18 | 0.00729 |
| 19 | (0, 1, 0, 0, 1) | 24 | 21 | 18 | 0.00729 |
| 20 | (1, 1, 0, 0, 1) | 16 | 14 | 12 | 0.00081 |
| 21 | (0, 0, 1, 0, 1) | 24 | 21 | 18 | 0.00729 |
| 22 | (1, 0, 1, 0, 1) | 16 | 14 | 12 | 0.00081 |
| 23 | (0, 1, 1, 0, 1) | 16 | 14 | 12 | 0.00081 |
| 24 | (1, 1, 1, 0, 1) | 8 | 7 | 6 | 0.00009 |
| 25 | (0, 0, 0, 1, 1) | 24 | 21 | 18 | 0.00729 |
| 26 | (1, 0, 0, 1, 1) | 16 | 14 | 12 | 0.00081 |
| 27 | (0, 1, 0, 1, 1) | 16 | 14 | 12 | 0.00081 |
| 28 | (1, 1, 0, 1, 1) | 8 | 7 | 6 | 0.00009 |
| 29 | (0, 0, 1, 1, 1) | 16 | 14 | 12 | 0.00081 |
| 30 | (1, 0, 1, 1, 1) | 8 | 7 | 6 | 0.00009 |
| 31 | (0, 1, 1, 1, 1) | 8 | 7 | 6 | 0.00009 |
| 32 | (1, 1, 1, 1, 1) | 0 | 0 | 0 | 0.00001 |
| | SUM | 640 | 560 | 480 | 1 |

**3.2 PLSA**

The reason for the high flexibility of the BAT to make-to-fit for various problems is in STEP B1 described in Section 3.1. We can calculate the function value of any new $X$ found in STEP B1 based on the problem objective. In the network reliability problem, our goal is to calculate the network reliability, and the reliability function $R(X)$ of each $X$ is

$$R(X) = \begin{cases} \Pr(X) & \text{if } X \text{ is connected} \\ 0 & \text{otherwise} \end{cases}. \quad (1)$$

From Eq. (1), we need to verify the connectivity of each vector $X$ obtained in STEP B1 from the BAT. To achieve this target efficiently, the PLSA was employed [28, 30]. The PLSA, first proposed in [28], originated from the layered-search algorithm (LSA) proposed in [37] to find $d$ simple paths, called $d$-MPs in [37], in acyclic networks. Owing to the efficiency and simplicity of the LSA, different versions of LSAs, the PLSA [37], GLSA [39], and TLSA [41], were proposed to validate the connectivity between two nodes, many nodes, and all nodes, respectively. Hence, the PLSA was employed to validate weather nodes 1, and $n$ was connected by at least one path in this study.

The pseudocode for the PLSA is presented below.

**Algorithm: PLSA**

**Input:** A state vector $X$.

**Output:** $X$ is connected or disconnected in $G(X)$.

**STEP P0.** Initialize the first layer $L_1 = \{1\}$, the index of the current layer $i = 2$, and $L_2 = \emptyset$.

**STEP P1.** Let $L_i = \{ v \in V \mid \text{if } X(e_{u,v}) = 1, u \in L_{i-1}, \text{ and } v \notin L_{i-1} \text{ in } G(X)\}$.

**STEP P2.** If $n \in L_i$, halt and $X$ is connected.

**STEP P3.** If $L_i = \emptyset$, halt and $X$ is disconnected. Otherwise, let $i = i + 1$, $L_i = \emptyset$, and go to STEP P1.

Based on the simple concept that nodes 1 and $n$ are connected if the current layer contains the node $n$ proposed in the LSA [37], the PLSA can determine whether $X$ is connected by validating whether nodes 1 and $n$ are connected in $G(X)$. Because each layer has at least one node and there are $n$ nodes, there are at most $n$ layers. Hence, the time complexity of PLSA is only $O(n)$.

For example, let $X = (1, 1, 1, 1, 1)$ in **Figure 1**, that is, $G(X) = G$. Based on the PLSA, we can verify the connectivity of $X$ using the procedure listed in Table 3.

**Table 3.** Verify the connectivity of $X$ using the PLSA.

| $i$ | $L_i$ | Remark |
|---|---|---|
| 1 | {1} | |
| 2 | {2, 3} | |
| 3 | {4} | $X$ is connected |

## 3.3 MCS and its important statistical characteristics

The MCS is a powerful and popular modeling method used to obtain approximate solutions for highly nonlinear, high-dimensional, and very complex problems [1-11]. Many MCS have been proposed, and the crude MCS implemented in [24, 25] was used here to calculate the approximate binary-state network reliability because of its simplicity and popularity. The MCS pseudocode is provided as follows [24, 25]:

**Algorithm: MCS**

**Input:** $G(V, E, \mathbf{D})$, the source node 1, the sink node $n$, and the number of simulations $N_{sim}$.

**Output:** The approximated reliability $R_{MCS}$.

**STEP M0.** Let $sim = j = 1$ and $N_{pass} = 0$.

**STEP M1.** If $\rho < \Pr(a_j)$, where $\rho$ is a random number generated within [0, 1] uniformly, let $X(a_j) = 1$; otherwise, $X(a_j) = 0$.

**STEP M2.** If $j < m$, let $j = j + 1$ and go to STEP M1.

**STEP M3.** If $X$ is connected, let $N_{pass} = N_{pass} + 1$.

**STEP M4.** If $sim < N_{sim}$, let $sim = sim + 1, j = 1$, and go to STEP M1.

**STEP M5.** Let $R_{MCS} = N_{pass}/N_{sim}$ and halt.

Because

$$E[\sum_{j=1}^{N_{sim}} \frac{R_{MCS,j}}{N_{sim}}] = R \tag{2}$$

$$Var[R_{MCS}] = \sum_{j=1}^{N_{sim}} \frac{R(1-R)}{N_{sim}}, \tag{3}$$

$R_{MCS}$ is an unbiased and consistent estimator of $R$ [24, 25].

In addition, the total number of MCS simulations $N_{sim}$ satisfies the following equation, which can be used to obtain the relative error $\varepsilon$ in the confidence interval $(1-\alpha)\%$ [24, 25]:

$$\frac{Z_{\alpha/2}^2}{4\varepsilon^2} \leq N_{sim}. \tag{4}$$

Eqs. (2)–(4) are important statistical characteristics of MCS, and their proofs can be obtained in textbooks on the MCS [29, 30]. Let $N_{sim} = 8$. The MCS is explained using the binary-state network shown in **Figure 1,** and its state distribution is listed in **Table 1**. The results are listed in **Table 4**: $X_2 = X_8$ for $N_{sim} = 8$. Hence, the MCS may have many duplications if $N_{sim}$ is increased and may not have sufficient number of events or vectors to realistically model problems.

**Table 4.** MCS Procedure for **Figure 1** based on **Table 1**.

| $i$ | $\rho_1$ | $X_i(a_1)$ | $\rho_2$ | $X_i(a_2)$ | $\rho_3$ | $X_i(a_3)$ | $\rho_4$ | $X_i(a_4)$ | $\rho_5$ | $X_i(a_5)$ | $X_i$ | Connected? | $N_{pass}$ |
|---|---|---|---|---|---|---|---|---|---|---|---|---|---|

| 1 | 0.16861 | 1 | 0.00281 | 1 | 0.48407 | 0 | 0.78768 | 0 | 0.26430 | 1 | (1, 1, 0, 0, 1) | Y | 1 |
|---|---|---|---|---|---|---|---|---|---|---|---|---|---|
| 2 | 0.24745 | 1 | 0.57434 | 1 | 0.94997 | 1 | 0.25461 | 1 | 0.81555 | 0 | (1, 1, 1, 1, 0) | Y | 2 |
| 3 | 0.37612 | 1 | 0.17782 | 1 | 0.69576 | 1 | 0.25661 | 1 | 0.95894 | 0 | (1, 1, 1, 1, 0) | Y | 3 |
| 4 | 0.99637 | 0 | 0.13337 | 1 | 0.93463 | 0 | 0.91058 | 0 | 0.83461 | 0 | (0, 1, 0, 0, 0) | N |   |
| 5 | 0.59383 | 1 | 0.08648 | 1 | 0.52618 | 1 | 0.21799 | 1 | 0.80758 | 0 | (1, 1, 1, 1, 1) | Y | 4 |
| 6 | 0.10244 | 1 | 0.41495 | 1 | 0.68811 | 1 | 0.64513 | 0 | 0.44071 | 1 | (1, 1, 1, 0, 1) | Y | 5 |
| 7 | 0.00730 | 1 | 0.25565 | 1 | 0.79960 | 0 | 0.82404 | 0 | 0.95794 | 0 | (1, 1, 0, 0, 0) | N |   |
| 8 | 0.46081 | 1 | 0.51113 | 1 | 0.09275 | 1 | 0.93567 | 0 | 0.12821 | 1 | (1, 1, 1, 0, 0) | N |   |
| SUM |   |   |   |   |   |   |   |   |   |   |   |   | 5/8=0.625 |

## 4. PROPOSED SELF-ADAPTIVE BAT-MCS

The proposed self-adaptive BAT-MCS, including the self-adaptive simulation-number algorithm to adjust the simulation number based on the importance of each super vector obtained from the BAT, the lower and upper bounds of each super vector used to reduce the number of simulations, the pseudocode to combine self-adaptive BAT and MCS, and the statistical characteristics of the proposed self-adaptive BAT-MCS are discussed in Section 4.

### 4.1 Super Family and Basic Idea of the Proposed Self-Adaptive BAT-MCS

The super vector $S$ is a special sub-vector, the value of whose $i$th coordinate denotes the state of the $i$th arc for $i = 1, 2, \ldots, \delta$ and $\|S\| = \delta \leq m$ [30]. For example, $S = (0, 0, 0)$ is a super set. The super vector was first proposed in the quick BAT [30] to skip disconnected vectors and verify their connectivity to reduce the run time in the BAT.

The super family is a novel concept and the core of the proposed self-adaptive BAT-MCS. A super family of the super vector $S$ is denoted by $\Omega_\delta(S) = \{\text{vector } X \mid X(a_i) = S(a_i) \text{ for } i = 1, 2, \ldots, \|S\| = \delta \text{ and } \|X\| = m\}$. For example, $\Omega_3(S) = \{(0, 0, 0, 0, 0), (0, 0, 0, 1, 0), (0, 0, 0, 0, 1), (0, 0, 0, 1, 1)\}$ if $S = (0, 0, 0)$ in **Figure 1**.

$R(\Omega_\delta(S))$ is the reliability of $\Omega_\delta(S)$ and

$$R(\Omega_\delta(S)) = \sum_{X \in \Omega_\delta(S)} R(X). \tag{5}$$

Because

$$R(X) = \begin{cases} \Pr(X) & \text{if } X \text{ is connected} \\ 0 & \text{otherwise} \end{cases}, \tag{6}$$

$$R(G) = \sum_{X \text{ is connected}} \Pr(X) = \sum_{\forall X \in \Omega} R(X), \tag{7}$$

$$\Omega_\delta = \bigcup_{\forall S} \Omega_\delta(S) \text{ and } \Omega_\delta(S) \cap \Omega_\delta(S^\#) = \varnothing \text{ if } S \neq S^\#, \tag{8}$$

we can rewrite Eq. (5) as Eq. (9).

$$R(\Omega_\delta(S)) = \sum_{X \in \Omega_\delta(S) \text{ and } X \text{ is connected}} \Pr(X) \tag{9}$$

Also, from Eq. (8) and Eq. (9) the network reliability $R(G)$ can be redefined as in Eq. (10):

$$R(G) = \sum_{X \in \Omega \text{ and } X \text{ is connected}} \Pr(X) = \sum_{\forall S} R(\Omega_\delta(S)) = R(\Omega_\delta). \tag{10}$$

The proposed BAT-MCS aims to efficiently solve Eq. (10) efficiently. Because

$$\Pr(S) = \prod_{i=1}^{\delta} \Pr(S(a_i)), \tag{11}$$

and the factoring theorem [42] gives

$$R(G) = \Pr(a)R(G(\mathbf{D}_{\Pr(a)=1})) + (1 - \Pr(a))R(G(\mathbf{D}_{\Pr(a)=0})), \tag{12}$$

Eq. (9) can be rewritten as in Eq. (11):

$$R(\Omega_\delta(S)) = R(G(\mathbf{D}_S)) \times \left[\prod_{i=1}^{\delta} \Pr(S(a_i))\right] = \Pr(S)R(G(\mathbf{D}_S)), \tag{13}$$

where

$$\mathbf{D}_S(a_j) = \begin{cases} \mathbf{D}(a_j) & \text{if } j > \|S\| \\ 1 & \text{if } j \leq \|S\| \text{ and } S(a_j) = 1. \\ 0 & \text{otherwsie} \end{cases} \tag{14}$$

It is still NP-hard to calculate $R(G(\mathbf{D}_S))$ in Eq. (13). Hence, the basic idea of the proposed algorithm is to use BAT to find all super vectors and the MCS to calculate the approximate reliability of $R(G(\mathbf{D}_S))$ corresponding to $\Omega_{\|S\|}(S)$ for each super vector $S$.

### 4.2 Lower and Upper Bounds of Super Vectors

Let $\|S\| = \delta$. The lower bound $L(S)$ of the super vector $S$ is a complete vector such that $L(S(a_i)) = S(a_i)$ for $i = 1, 2, \ldots, \delta$ and $L(S(a_i)) = 0$ for $i = (\delta+1), (\delta+2), \ldots, m$. $U(S(a_i)) = S(a_i)$ for $i = 1, 2, \ldots,$ $\delta$ and $U(S(a_i)) = 1$ for $i = (\delta+1), (\delta+2), \ldots, m$, where $U(S)$ is the upper bound of the super vector $S$. For example, $L(S) = (0, 0, 0, 0, 0)$ and $U(S) = (0, 0, 0, 1, 1)$ if $S = (0, 0, 0)$.

If $L(S)$ is connected, any vector $X$ is also connected because $L(S) \leq X$ for all $X \in \Omega(S)$. Similarly, if $U(S)$ is disconnected, any vector $X$ is also disconnected because $X \leq U(S)$. However, $S$ is neither disconnected nor connected if $L(S)$ is disconnected or $U(S)$ is connected.

Let $L_\delta$, $U_\delta$, and $F_\delta$ be super families such that $L(S) \in L_\delta$ is connected, $U(S) \in U_\delta$ be disconnected, and $F_\delta = \Omega_\delta - (L_\delta \cup U_\delta)$, for all super vectors $S \in \Omega_\delta$. From the properties of $L_\delta$, $U_\delta$, $F_\delta$, and Eq. (13), we have

$$R(U_\delta) = 0, \tag{15}$$

$$R(L_\delta) = \sum_{S \in L_\delta} \prod_{i=1}^{\delta} \Pr(S(a_i)) = \sum_{S \in L_\delta} \Pr(S), \tag{16}$$

$$R(F_\delta) = \sum_{S \in F_\delta} \Pr(S) R(G(\mathbf{D}_S)). \tag{17}$$

Because

$$R(G) = R(\Omega_\delta) = R(L_\delta) + R(U_\delta) + R(F_\delta), \tag{18}$$

based on Eqs.(15)–(17), we have

$$R(G) = \sum_{S \in L_s} \Pr(S) + \sum_{S \in F_\delta} \Pr(S) R(G(\mathbf{D}_S)). \tag{19}$$

For example, in **Figure 1**, $X \in \Omega(S_u)$ is disconnected, where $S_u = (0, 0)$, because $U(S_u) = (0, 0, 1, 1, 1)$ is disconnected. Hence, $S_u \in U_2$, and there is no need to find all the vectors in $\Omega(S_u)$ to calculate the exact reliability, that is, $\Pr(\Omega(S_u)) = 0$. In the same way, $X \in \Omega(S_l)$ is connected because $L(S_l) = (1, 0, 0, 1, 0)$ is connected if $S_l = (1, 0, 0, 1)$. Hence, $\Pr(S_l) = 0.9 \times 0.2 \times 0.3 \times 0.6 = 0.0324 = \Pr(\Omega(S_l))$ from **Table 1**.

As another example, $S_f = (0, 1, 0)$ is neither disconnected nor connected because $L(S_f) = (0, 1, 0, 0, 0)$ is disconnected and $U(S_f) = (0, 1, 0, 1, 1)$ is connected. Hence, $S_u \in U_2 \subseteq \Omega_2$, $S_l \in L_4 \subseteq \Omega_4$, and $S_l \in F_3 \subseteq \Omega_3$.

The MCS in the proposed BAT-MCS was used to estimate $R(G(\mathbf{D}_S))$ in Eq. (19). Let

$R_{\text{MCS}}(G(\mathbf{D}_S))$ be the estimator of $R(G(\mathbf{D}_S))$ obtained from the MCS. Hence, the estimator of $R(G)$ in Eq. (19) can be written as:

$$R_{\text{BAT-MCS}}(G) = \sum_{S \in L_s} \Pr(S) + \sum_{S \in F_\delta} \Pr(S) R_{\text{MCS}}(G(\mathbf{D}_S)). \tag{20}$$

For example, if $S = (1, 0)$ in **Figure 1**, we have $\Pr(S) = p_1 \times q_2 = 0.9 \times 0.2 = 0.18$, based on **Table 1**. $G(\mathbf{D}_S)$ is depicted in **Figure 3,** and $\mathbf{D}_S$ is listed in **Table 5**, where the probabilities of the doubled line $a_1$ and dashed line $a_2$ are one and zero, respectively. If $R_{\text{MCS}}(G(\mathbf{D}_S)) = 0.2$ after implementing the MCS, then $R_{\text{BAT-MCS}}(\Omega(S_i)) = \Pr(S_i) \times R_{\text{MCS}}(G(\mathbf{D}_S)) = 0.18 \times 0.2 = 0.036$.

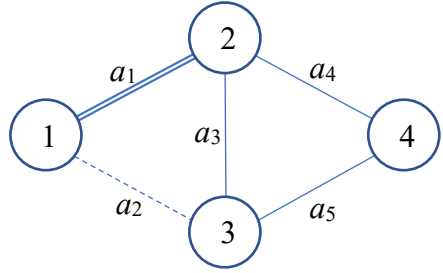

**Figure 3.** New network $G_i$.

**Table 5.** Binary-state distribution $\mathbf{D}_S$ for **Figure 3**.

| $i$ | $\Pr(a_i)$ |
|---|---|
| 1 | 1.0 |
| 2 | 0 |
| 3 | 0.7 |
| 4 | 0.6 |
| 5 | 0.5 |

**4.3 Self-Adaptive Simulation-Number Algorithm**

The number of multiple simulation replications is a critical factor determining MCS performance, including the quality and variance of solutions and efficiency [24, 25]. More than millions of random variates may be required in one simulation to obtain an accurate estimator [43]. Hence, to reduce variance and maximize the probability of correct selection, it is crucial to ensure that these critical events have more simulation numbers than less important events.

A novel self-adaptive algorithm is proposed to self-adaptively allocate simulation numbers to different events based on their importance.

Let $N_{sim}$ be the total number of simulations, and $n_{sim}(S)$ be the number of simulations of $R_{MCS}(G(\mathbf{D}_S))$ for the super vector $S$. It is not necessary to use MCS to approximate the probabilities of these super vectors in $U_\delta$ and $L_\delta$ from Eq. (20) in Section 4.2. Hence,

$$N_{sim} = \sum_{S \in \Omega_\delta} n_{sim}(S) = \sum_{S \in F_\delta} n_{sim}(S). \tag{21}$$

Eq. (21) implements the MCS on these super vectors in $F_\delta$ only. To further improve the method of allocating the number of simulations, the proposed self-adaptive algorithm allocates the simulation number based on the probability of each super vector in $F_\delta$ as follows:

$$n_{sim}(S) = \begin{cases} n^*_{sim}(S) + N_{sim} - \sum_{S \in F_s} n^*_{sim}(S) & \text{if } \Pr(S^*) \le \Pr(S) \text{ for all } S^* \in F_\delta \\ n^*_{sim}(S) & \text{otherwise} \end{cases}, \tag{22}$$

where

$$n^*_{sim}(S) = \left\lfloor N_{sim} \times \frac{\Pr(S)}{\sum_{S \in F_s} \Pr(S)} \right\rfloor. \tag{23}$$

Eq. (23) can be used to allocate the number of simulations for critical and non-critical events. For example, let $N_{sim} = 1024$ and $\delta = 2$ in **Figure 1**. Based on the BAT, we have $\Omega_2 = \{(0, 0), (1, 0), (0, 1), \text{and } (1, 1)\}$. From **Table 1**, Eq. (22), and Eqs. (21)–(23), we obtain $n_{sim}(S)$ for all super vectors $S \in F_\delta$ and $\delta = 2$, as shown in **Table 6**, where $\Pr^{\#}(S_i) = \Pr(S_i) / [\Pr(S_2) + \Pr(S_3) + \Pr(S_4)]$.

Table 6. Calculated $n_{sim}(S)$ for all $S \in \Omega_2$ based on **Figure 1**, **Table 1**, and Eq. (22).

| i | $S_i(a_1)$ | $S_i(a_1)$ | $\Pr(S_i(a_1))$ | $\Pr(S_i(a_2))$ | $\Pr(S_i)$ | $\Pr^{\#}(S_i)$ | $N_{sim} \times \Pr^{\#}(S_i)$ | $n^*_{sim}(S_i)$ | $n_{sim}(S_i)$ |
|---|---|---|---|---|---|---|---|---|---|
| 1 | 0 | 0 | | | | | | | |
| 2 | 1 | 0 | 0.9 | 0.2 | 0.18 | 0.183673 | 188.0816 | 188 | 188 |
| 3 | 0 | 1 | 0.1 | 0.8 | 0.08 | 0.081633 | 83.59184 | 83 | 83 |
| 4 | 1 | 1 | 0.9 | 0.8 | 0.72 | 0.734694 | 752.3265 | 752 | 753 |
| SUM | | | | | 0.98 | 1 | 1024 | 1023 | 1024 |

## 4.4 Statistical Characteristics of the Self-Adaptive BAT-MCS

The statistical characteristics of the proposed self-adaptive BAT-MCS are discussed in this section. Eq. (2) and Eq. (20) imply that the estimator obtained from the self-adaptive BAT-MCS is also unbiased and consistent with that obtained from the traditional MCS. Hence, we focus only on the variance because it is a measure of the quality and robustness of the obtained estimators.

If $\delta = 0$, we have $\Omega_\delta = \varnothing$, and the BAT-MCS is exactly the same as the MCS. In contrast, if $\delta = m$, then the BAT-MCS is the same as the BAT. We assume that in the proposed self-adaptive BAT-MCS, the number of coordinates of each super vector is $\delta = 1, 2, \ldots, (m-1)$, and there are $\omega = 2^\delta$ super vectors in total excluding the super vectors in $U_\delta$ and $L_\delta$. $\lambda$ simulations are required for each super vector, that is, the total number of simulations is $\lambda\omega$. To ensure a fair comparison between the estimators obtained from the BAT-MCS and the MCs, let the number of simulations of both algorithms be $N_{sim}$, that is, $\lambda\omega = N_{sim}$, and that the self-adaptive simulation-number algorithm is not used.

Using Eq. (3) and $\lambda\omega = N_{sim}$, the variance of the estimators obtained from the traditional MCS can be rewritten as Eq. (24).

$$Var[R_{MCS}] = \frac{R \times (1-R)}{N_{sim}} = \frac{R - R^2}{\omega\lambda} \tag{24}$$

Further, from Eq. (3) and Eq. (20), the variance of the estimators obtained from the self-adaptive BAT-MCS is as follows.

$$Var[R_{BAT\text{-}MCS}] = \sum_{j=1}^{\omega} \frac{\Pr^2(S_i)\left[R(G(\mathbf{D}_{S_i})) \times (1 - R(G(\mathbf{D}_{S_i})))\right]}{\lambda}$$

$$= \sum_{j=1}^{\omega} \frac{R \times (\Pr(S_i) - R)}{\lambda}$$

$$= \frac{R \times \sum_{j=1}^{\omega} \Pr(S_i) - \omega R^2}{\lambda}$$

$$\leq \frac{R - \omega R^2}{\lambda}. \tag{25}$$

Because $\delta \geq 1$ and $\omega = 2^\delta$, we have $\omega \geq 2$ and

$$Var[R_{MCS}] - Var[R_{BAT\text{-}MCS}] = \frac{R}{\lambda}\left[R(\omega - \frac{1}{\omega}) + (\frac{1}{\omega} - 1)\right]$$

$$= \frac{(\omega-1)R}{\omega\lambda}[R(\omega+1)-1]. \quad (26)$$

Hence,

$$Var[R_{MCS}] - Var[R_{BAT\text{-}MCS}] \geq 0 \Leftrightarrow R \geq \frac{1}{\omega+1} = \frac{1}{2^\delta+1}. \quad (27)$$

Eq. (27) implies that the variance of the estimator obtained from the BAT-MCS is less than that obtained from the MCS if $R \geq \frac{1}{2^\delta+1}$. Hence, the proposed BAT-MCS is more robust than the traditional MCS.

**4.5 Pseudocode and Time Complexity**

The pseudocode for calculating the approximate reliability of binary-state networks using the proposed self-adaptive BAT-MCS by integrating the lower and upper bounds of super vectors presented in Section 4.1 and the self-adaptive simulation-number algorithm discussed in Section 4.2 is presented here.

**Algorithm: BAT-MCS**

**Input:** A binary-state network $G(V, E, \mathbf{D})$, the source node 1, and the sink node $n$.

**Output:** The approximate reliability $R_{BAT\text{-}MCS}$ under pregiven values $\delta$, $\Pr(a)$ for all $a \in E$, and $N_{sim}$.

**STEP 0.** Let $\delta$-tuple super vector $X$ be a vector zero, $i = 1$, $k = R_{BAT\text{-}MCS} = 0$, and go to STEP 3.

**STEP 1.** If $X(a_i) = 0$, let $X(a_i) = 1$, $i = 1$, and go to STEP 3.

**STEP 2.** If $i < \delta$, let $X(a_i) = 0$, $i = i + 1$, and go to STEP 1. Otherwise, go to STEP 6.

**STEP 3.** If $L(X)$ is connected, let $R_{BAT\text{-}MCS} = R_{BAT\text{-}MCS} + \Pr(X)$ and go to STEP 1.

**STEP 4.** If $U(X)$ is disconnected, go to STEP 1.

**STEP 5.** Let $k = k + 1$, $S_k = X$, and go to STEP 1.

**STEP 6.** Calculate $n_{sim}(S_i)$ based on Eq. (22) for $i = 1, 2, \ldots, k$ and let $i = 1$.

**STEP 7.** Let $j = \delta + 1$ and $N_{pass} = sim = 0$.

**STEP 8.** If $\rho < \Pr(a_j)$, where $\rho$ is a random number generated within [0, 1] uniformly, let $S_i(a_j) = 1$; otherwise, $S_i(a_j) = 0$.

**STEP 9.** If $j < m$, let $j = j + 1$ and go to STEP 8.

**STEP 10.** If $S_i$ is connected, let $N_{pass} = N_{pass} + 1$.

**STEP 11.** If $sim < n_{sim}(S_i)$, let $sim = sim + 1$, $j = \delta + 1$, and go to STEP 8.

**STEP 12.** Let $R_{BAT\text{-}MCS} = R_{BAT\text{-}MCS} + Pr(S_i) \times N_{pass}/N_{sim}$.

**STEP 13.** If $i < k$, let $i = i + 1$ and go to STEP 7. Otherwise, halt.

The first loop from STEP 0 to STEP 5 is based on the BAT and used to find the super vectors. The second loop from STEP 6 to STEP 13 is mainly based on the MCS and used to calculate the reliability of each feasible super vector with the proposed self-adaptive simulation-number algorithm.

STEP 0 initializes the super vector to the $\delta$-tuple vector zero, the index of the current coordinate to $i = 1$, and the approximate reliability to $R_{BAT\text{-}MCS} = 0$. STEPs 1 and 2 comprise the BAT for finding each super vector with $\delta$ coordinates. If the value of the current coordinate $X(a_i)$ is zero, then $i$ is reset to one and a new super vector $X$ is found.

The lower and upper bounds of a new vector are required to verify its connectivity using PLSA in STEPs 3 and 4. STEP 3 calculates $Pr(X)$ if $X \in L_\delta$; STEP 4 abandons $X$ if $X \in U_\delta$, and STEP 5 adds $X$ to $F_\delta$ if $X \notin (L_\delta \cup U_\delta)$.

STEP 6 implements the proposed self-adaptive algorithm to determine the simulation number of such a super vector based on the probability of each super vector stored in STEP 5.

STEPs 7–9 simulates the value of the $j$th coordinate of each super vector $S_i$ in the $(sim)^{th}$ simulation for $j = (\delta+1), (\delta+2), \ldots, m$, $sim = 1, 2, \ldots, n_{sim}(S_i)$, and $i = 1, 2, \ldots, k$. STEP 10 determines whether the $(sim)^{th}$ simulation was successful by using the PLSA to verify the connectivity of the related vector. If it is connected, the value of $N_{pass}$ increases by one. STEP 11 tests whether the current simulation is the last one. STEP 12 adds $Pr(S_i) \times N_{pass}/N_{sim}$ to $R_{BAT\text{-}MCS}$. STEP 12 also enforces the stopping criterion by checking whether the current super vector is the last one.

The number of super vectors found is $2^\delta$ because the number of their coordinates are all equal to $\delta$ and the value of each coordinate, that is, $X(a_i)$ for $i = 1, 2, \ldots, \delta$, is either 0 or 1. Hence, the total time complexity is $O(n2^\delta + 2^{(\delta+1)}) = O(n2^\delta)$ for STEPs 1 and 2, where $O(n_\delta)$ results from the PLSA [37] and $O(2^{\delta+1})$ from the BAT [39].

After all the super vectors in $F_\delta$ are found, the MCS is implemented for each of them. The number of MCS simulations is $N_{sim}$, and each simulation needs to generate the values of $(m - \delta)$ coordinates, that is, $S_i(a_j)$ for $j = (\delta + 1), (\delta + 2), …, m$. Additionally, the PLSA is implemented to verify the connectivity of $S_i$ after extending it from a $\delta$-tuple to an $m$-tuple vector for each simulation [28]. Hence, the time complexity is $O((m - \delta + n) N_{sim})$, where $O(n)$ is the time complexity required to implement the PLSA [28].

Thus, the total complexity is $O(n2^\delta + [(m - \delta + n) N_{sim}]2^\delta) = O([n + (m - \delta + n) N_{sim}]2^\delta)$. If no MCS is implemented in BAT-MCS, that is, $\delta = m$ and $N_{sim} = 0$, then we have $O([n + (m-\delta+n)N_{sim}]2^\delta) = O(n\,2^m)$, which is also the time complexity for implementing the BAT to calculate the binary-state network reliability [39]. In contrast, if no BAT is applied in BAT-MCS, that is, $\delta = 0$, then we have $O([n + (m-\delta+n) N_{sim}]2^\delta) = O((m + n) N_{sim})$, which is the time complexity for implementing the MCS.

Thus, in terms of the time complexity, the efficiency of the proposed BAT-MCS is higher than that of the BAT if $\delta << m$ and very close to that of the MCS if $\delta$ is very small.

## 5 BENCHMARK EXAMPLE AND EXPERIMENTS ON A LARGE-SCALE NETWORK

Section 5 presents the step-by-step implementation of the proposed self-adaptive BAT-MCS based on a simple benchmark network. A computational experiment on a large-scale example is also provided to analyze the relationships among $R_{MCS}$, $R_{BAT-MCS}$, the number of coordinates $\delta$ in super vectors, $n_{sim}(S)$ for all super vectors $S$, and $\Pr(a)$ for all $a \in E$.

### 5.1 Step-By-Step BAT-MCS Procedure

As mentioned in Section 1, calculating the binary-state network reliability is both NP-hard and #P-hard [17, 18]. The simplest and most cited benchmark network in network reliability research is depicted in **Figure 1** [17, 18, 28, 39]. To help readers realize the proposed self-adaptive BAT-MCS easily, the self-adaptive BAT-MCS is tested, as shown in **Figure 1,** to exemplify the procedure of the self-adaptive BAT-MCS for calculating the approximate binary-state network reliability.

The details of the step-by-step self-adaptive BAT-MCS procedure to calculate $R_{\text{BAT-MCS}}$ without STEP 6 are provided here, where $\delta = 2$, $n_{\text{sim}}(S) = N_{\text{sim}} = 2$ for all super vectors $S$, and $\Pr(a_i)$ is as provided in **Table 1** for $i = 1, 2, \ldots, 5$:

**Solution:**

**STEP 0.** Let $X = (0, 0)$, $i = 1$, $j = \delta + 1 = 3$, $N_{\text{pass}} = R_{\text{BAT-MCS}} = 0$, and go to STEP 3.

**STEP 3.** Because $L(X) = (0, 0, 0, 0, 0)$ is disconnected, go to STEP 4.

**STEP 4.** Because $U(X) = (0, 0, 0, 1, 1)$ is disconnected, go to STEP 1.

**STEP 1.** Because $X(a_1) = 0$, let $X(a_1) = 1$, $i = 1$, and go to STEP 3. Note that $X = (1, 0)$ now.

$$\vdots$$

**STEP 5.** Let $k = k + 1 = 3$, $S_3 = X = (1, 1)$, and go to STEP 1.

**STEP 1.** Because $X(a_1) = 1$, go to STEP 2.

**STEP 2.** Because $i = 1 < \delta = 2$, let $X(a_1) = 0$, $i = i + 1 = 2$, and go to STEP 1.

**STEP 1.** Because $X(a_2) = 1$, go to STEP 2.

**STEP 2.** Because $i = \delta = 2$, go to STEP 7. Note that the part for the self-adaptive simulation-number algorithm, i.e., STEP 6, is skipped.

**STEP 7.** Let $j = \delta + 1 = 3$ and $N_{\text{pass}} = \text{sim} = 0$.

**STEP 8.** Assume that $\rho = 0.93464$ generated within $[0, 1]$ uniformly. Because $\Pr(a_3) = 0.7 < \rho$, let $X(a_3) = 0$.

**STEP 9.** Because $j = 3 < m = 5$, let $j = j + 1 = 4$ and go to STEP 3.

**STEP 8.** Generate a random number, say $\rho = 0.91058$, within $[0, 1]$ uniformly. Because $\Pr(a_4) = 0.6 < \rho$, let $X(a_4) = 0$.

**STEP 9.** Because $j = 4 < m = 5$, let $j = j + 1 = 5$ and go to STEP 3.

**STEP 3.** Suppose that $\rho = 0.83461$ generate within $[0, 1]$ uniformly. Because $\rho < \Pr(a_5) = 0.5$, let $X(a_5) = 0$.

**STEP 4.**  Because $j = 5 = m$, go to STEP 5.

**STEP 5.**  Because $X = (1, 0, 0, 0, 0)$ is disconnected, go to STEP 6.

**STEP 6.**  Because sim $= 1 < N_{sim} = 2$, let sim $=$ sim $+ 1 = 2$, $j = \delta + 1 = 3$, and go to STEP 3.

**STEP 3.**  Suppose that $\rho = 0.52618$ generate within $[0, 1]$ uniformly. Because $\rho < \Pr(a_3) = 0.7$, let $X(a_3) = 1$.

**STEP 4.**  Because $j = 3 < m = 5$, let $j = j + 1 = 4$ and go to STEP 3.

**STEP 3.**  Generate a random number, say $\rho = 0.217987$, within $[0, 1]$ uniformly. Because $\rho < \Pr(a_4) = 0.6$, let $X(a_4) = 1$.

**STEP 4.**  Because $j = 4 < m = 5$, let $j = j + 1 = 5$ and go to STEP 3.

**STEP 3.**  Suppose that $\rho = 0.80758$ generate within $[0, 1]$ uniformly. Because $\Pr(a_5) = 0.5 < \rho$, let $X(a_5) = 0$.

**STEP 4.**  Because $j = 5 = m$, go to STEP 5.

**STEP 5.**  Because $X = (1, 0, 1, 1, 0)$ is connected, let $N_{pass} = N_{pass} + 1 = 1$.

**STEP 6.**  Because sim $= N_{sim} = 2$, go to STEP 7.

**STEP 7.**  Let $R_{\delta, N_{sim}} = R_{\delta, N_{sim}} + \Pr(X(1:2)) N_{pass}/N_{sim} = 0.018 \cdot 1/2 = 0.009$ and go to STEP 1, where $\Pr(X(1:2)) = \Pr(X(a_1)) \Pr(X(a_2)) = 0.9 \cdot 0.2 = 0.018$.

$$\vdots$$

The complete procedure is listed in **Table 7** and the final approximated reliability $R^* = 0.53$.

**Table 7.** BAT-MCS procedure for **Figure 1** and **Table 1**.

| c | X | Pr(X) | $\rho_3$ | $X(a_3)$ | $\rho_4$ | $X(a_4)$ | $\rho_5$ | $X(a_5)$ | X | Connected? | $R_{BAT\text{-}MCS}(X)$ |
|---|---|---|---|---|---|---|---|---|---|---|---|
| 1 | (0, 0) | | | | | | | | | | |
| 2 | (1, 0) | 0.18 | 0.69576 | 1 | 0.25661 | 1 | 0.95894 | 0 | (1, 0, 1, 1, 0) | Y | 0.18×(1/2) |
|   |        |      | 0.93463 | 0 | 0.91058 | 0 | 0.83461 | 0 | (1, 0, 0, 0, 0) | N | |
| 3 | (0, 1) | 0.08 | 0.52618 | 1 | 0.21799 | 1 | 0.80758 | 0 | (0, 1, 1, 1, 0) | Y | 0.08×(1/2) |
|   |        |      | 0.68811 | 1 | 0.64513 | 0 | 0.44071 | 1 | (0, 1, 1, 0, 1) | Y | 0.08×(1/2) |
| 4 | (1, 1) | 0.72 | 0.79960 | 0 | 0.82404 | 0 | 0.95794 | 0 | (1, 1, 0, 0, 0) | N | |
|   |        |      | 0.09275 | 1 | 0.93567 | 0 | 0.12821 | 1 | (1, 1, 1, 0, 1) | Y | 0.72×(1/2) |
| SUM | | | | | | | | | | | 0.53 |

The super vector $(0, 0)$ was removed from Table 7 because $(0, 0) \in U_2$. Hence, the proposed lower-bound and upper-bound methods reduce the number of super vectors. Comparing the results in Table 4 with those in Table 7, no duplications are observed using the proposed self-adaptive BAT-MCS. This is because there are no duplicate super vectors obtained from the BAT, and using super vectors separates the problem into disjoint subproblems to reduce the probability of duplications.

Hence, the proposed BAT-MCS can reduce the number of simulations, decrease the probability of duplications, and reduce the variance.

## 5.2 Computation Experiments on a Large-Scale Network

To investigate the practical performance of the proposed self-adaptive BAT-MCS and the effects of $\delta$, $N_{sim}$, and $Pr(a)$ on the performance of self-adaptive BAT-MCS, an extension test was conducted on a large-scale benchmark network, which is widely employed to verify the performance of new algorithms (**Figure 4**). In **Figure 4**, we have $n = 10$, $m = 21$, and the number of vectors is $2^m = 2{,}097{,}152$ [17, 18].

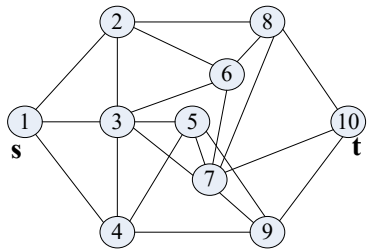

**Figure 4.** Large-scale benchmark binary-state networks.

The proposed self-adaptive BAT-MCS was programmed in DEV C$^{++}$ and executed on an Intel Core i7-10750H CPU @2.60 GHz with 64 GB RAM on a 64-bit Windows 11 system. The values of $\delta$ were 0, 1, 3, 5, …, 19; those of $N_{sim}$ were $2^{15}$, $2^{17}$, $2^{19}$, $2^{21}$, $2^{23}$; and those of $Pr(a)$ were 0.1, 0.3, 0.5, 0.7, 0.9 for all $a \in E$. Hence, a total of 21×5×5 = 525 tests were conducted. Each test was run $N_{run} = 30$ times, and the results are presented in **Table 8**, **Table 11**, and **Table 13**. In addition, the best values are shown in bold.

## 5.2.1 Simulation Number

**Table 8** shows that the value of $\|F_\delta\|/\|\Omega_\delta\|$ decreases with increasing $\delta$, e.g., $\|F_8\|/\|\Omega_8\| = 85.9375\%$ and $\|F_{20}\|/\|\Omega_{20}\| = 21.3581\%$. Hence, the benefit of using the lower and the upper bounds of the super vectors tends to increase from small $\delta$ to large $\delta$.

**Table 8.** Results for $\|\Omega_\delta\|$, $\|L_\delta\|$, $\|U_\delta\|$, $\|F_\delta\|$, $\|F_\delta\|/\|\Omega_\delta\|$ using the self-adaptive BAT-MCS.

| $\delta$ | $\|\Omega_\delta\|$ | $\|L_\delta\|$ | $\|U_\delta\|$ | $\|F_\delta\|$ | $\|F_\delta\|/\|\Omega_\delta\|$ |
|---|---|---|---|---|---|
| 0 | | | | | |
| 1 | 2 | 0 | 0 | 2 | 100.0000% |
| 2 | 4 | 0 | 0 | 4 | 100.0000% |
| 3 | 8 | 0 | 1 | 7 | 87.5000% |
| 4 | 16 | 0 | 2 | 14 | 87.5000% |
| 5 | 32 | 0 | 4 | 28 | 87.5000% |
| 6 | 64 | 0 | 9 | 55 | 85.9375% |
| 7 | 128 | 0 | 18 | 110 | 85.9375% |
| 8 | 256 | 0 | 36 | 220 | 85.9375% |
| 9 | 512 | 0 | 72 | 440 | 85.9375% |
| 10 | 1024 | 0 | 152 | 872 | 85.1563% |
| 11 | 2048 | 0 | 304 | 1744 | 85.1563% |
| 12 | 4096 | 0 | 710 | 3386 | 82.6660% |
| 13 | 8192 | 0 | 1420 | 6772 | 82.6660% |
| 14 | 16384 | 0 | 2996 | 13388 | 81.7139% |
| 15 | 32768 | 0 | 5992 | 26776 | 81.7139% |
| 16 | 65536 | 0 | 12834 | 52702 | 80.4169% |
| 17 | 131072 | 0 | 25668 | 105404 | 80.4169% |
| 18 | 262144 | 0 | 51336 | 210808 | 80.4169% |
| 19 | 524288 | 193351 | 106981 | 223956 | 42.7162% |
| 20 | 1048576 | 570388 | 254232 | 223956 | **21.3581%** |

**Table 9.** $n_{sim}(S)$ without using the proposed self-adaptive simulation algorithm in **Figure 4**.

| $\delta$ | $2^{11}$ | $2^{13}$ | $2^{15}$ | $2^{17}$ | $2^{19}$ |
|---|---|---|---|---|---|
| 0 | 2048 | 8192 | 32768 | 131072 | 524288 |
| 1 | 1024 | 4096 | 16384 | 65536 | 262144 |
| 3 | 256 | 1024 | 4096 | 16384 | 65536 |
| 5 | 64 | 256 | 1024 | 4096 | 16384 |
| 7 | 16 | 64 | 256 | 1024 | 4096 |
| 9 | 4 | 16 | 64 | 256 | 1024 |
| 11 | 1 | 4 | 16 | 64 | 256 |
| 13 | 0.25 | 1 | 4 | 16 | 64 |
| 15 | 0.0625 | 0.25 | 1 | 4 | 16 |
| 17 | 0.015625 | 0.0625 | 0.25 | 1 | 4 |
| 19 | 0.003906 | 0.015625 | 0.0625 | 0.25 | 1 |

**Table 9** shows that the value of $n_{sim}(S)$ decreases with the number of $\delta$. Without using the proposed self-adaptive simulation algorithm, each super vector $S$ needs to be simulated even if $S \notin F_{||S||}$. Hence, $n_{sim}(S) = N_{sim}/2^{\delta}$ can be less than one if $N_{sim} < 2^{\delta}$. Thus, it is difficult to increase the $\delta$ to improve the variance without using the proposed self-adaptive simulation algorithm.

**Table 10** shows that the average of $n_{sim}(S)$ increases after using the proposed self-adaptive simulation algorithm for each super vector. Hence, the proposed self-adaptive simulation algorithm can increase the $\delta$ to reduce the variance and improve the quality of the estimator obtained from the proposed self-adaptive BAT-MCS based on Eq. (25). However, **Table 10** also shows that the number of $n_{sim}(S)$ is reduced dramatically at larger $\delta$, for example, $n_{sim}(S) = 4 + 0.8387 = 4.8387$ for $\delta = 13$ and $N_{sim} = 2^{15}$, which results in the generation of an insufficient number of samples to represent all situations of the network and improve the solution quality and variance from the proposed self-adaptive simulation algorithm.

Hence, selecting an appropriate $\delta$ is the critical concern in the proposed algorithm. Note that the number of vectors is $2^m = 2,097,152$ only, that is, any MCS with more than 2,097,152 simulations can be replaced by the BAT.

**Table 10.** Increase in average $n_{sim}(S)$ with the proposed self-adaptive simulation-algorithm.

| $\delta$ | $\|F_\delta\|$ | $2^{11}$ | $2^{13}$ | $2^{15}$ | $2^{17}$ | $2^{19}$ |
|---|---|---|---|---|---|---|
| 0 | 0 | 0.0000 | 0.0000 | 0.0000 | 0.0000 | 0.0000 |
| 1 | 2 | 0.0000 | 0.0000 | 0.0000 | 0.0000 | 0.0000 |
| 3 | 7 | **36.5714** | **146.2857** | **585.1429** | **2340.5714** | **9362.2857** |
| 5 | 28 | 9.1429 | 36.5714 | 146.2857 | 585.1429 | 2340.5714 |
| 7 | 110 | 2.6182 | 10.4727 | 41.8909 | 167.5636 | 670.2545 |
| 9 | 440 | 0.6545 | 2.6182 | 10.4727 | 41.8909 | 167.5636 |
| 11 | 1744 | 0.1743 | 0.6972 | 2.7890 | 11.1560 | 44.6239 |
| 13 | 6772 | 0.0524 | 0.2097 | 0.8387 | 3.3550 | 13.4200 |
| 15 | 26776 | 0.0140 | 0.0559 | 0.2238 | 0.8951 | 3.5805 |
| 17 | 105404 | 0.0038 | 0.0152 | 0.0609 | 0.2435 | 0.9741 |
| 19 | 223956 | 0.0052 | 0.0210 | 0.0838 | 0.3353 | 1.3410 |

**5.2.2 Average Absolute Errors $\bar{\varepsilon}$**

**Table 11** shows that overall, both the absolute errors $\bar{\varepsilon}_{MCS} = |R - \frac{\sum_{i=1}^{N_{run}} R_{MCS,i}}{N_{run}}|$ and $\bar{\varepsilon}_{BAT-MCS} = |R - \frac{\sum_{i=1}^{N_{run}} R_{BAT-MCS,i}}{N_{run}}|$ increase with increasing $\Pr(a)$ up to a peak and then decrease. For example, for

$\delta = 1$, $\bar{\varepsilon}_{BAT-MCS}$ is 0.000047 at $p = 0.1$, reaches its peak $\bar{\varepsilon}_{BAT-MCS} = 0.000400$ at $p = 0.3$, and then decreases. The minimum average $\bar{\varepsilon} = 0.000765$ is obtained at $p = 0.1$, which may be because $p = \Pr(a) = 0.1$ is less than the other $p$ values and results in lower values of $R$ and $\bar{\varepsilon}$.

For $\delta = 1$, 3, and 5, the proposed algorithm has a better average $\bar{\varepsilon}$, after which $\bar{\varepsilon}_{MCS}$ is better than $\bar{\varepsilon}_{BAT-MCS}$. Hence, a smaller $\delta$ can improve $\bar{\varepsilon}_{BAT-MCS}$. This observation confirms that the critical concern in the proposed algorithm is to find the optimal $\delta$ to obtain a better result, as discussed at the end of Section 5.2.1.

Table 11. Results of $\bar{\varepsilon}$ based on $\delta$ and $p$.

| $\delta$ | 0.1 | 0.3 | 0.5 | 0.7 | 0.9 | avg |
|---|---|---|---|---|---|---|
| 0 | 0.000101 | 0.001059 | 0.000294 | 0.000487 | 0.000098 | 0.000408 |
| 1 | **0.000047** | **0.000400** | 0.000250 | 0.000294 | 0.000067 | **0.000212** |
| 3 | 0.000060 | 0.000404 | 0.000798 | **0.000177** | **0.000037** | 0.000295 |
| 5 | 0.000129 | **0.000198** | 0.000491 | 0.000238 | 0.000120 | 0.000235 |
| 7 | 0.000132 | 0.001055 | 0.000782 | 0.000247 | 0.001000 | 0.000643 |
| 9 | 0.000355 | 0.002324 | 0.000306 | 0.003404 | 0.003464 | 0.001971 |
| 11 | 0.000888 | 0.020265 | **0.000184** | 0.038573 | 0.022367 | 0.016455 |
| 13 | 0.001284 | 0.054808 | 0.130349 | 0.142927 | 0.035852 | 0.073044 |
| 15 | 0.001618 | 0.095817 | 0.260396 | 0.306397 | 0.052912 | 0.143428 |
| 17 | 0.002316 | 0.125947 | 0.390517 | 0.448035 | 0.085341 | 0.210431 |
| 19 | 0.001488 | 0.083787 | 0.225586 | 0.196869 | 0.023822 | 0.106310 |
| avg | **0.000765** | 0.035097 | 0.091814 | 0.103422 | 0.020462 | |

**Table 12** compares $\bar{\varepsilon}_{BAT-MCS}$ with $\bar{\varepsilon}_{MCS}$ based on $\delta$ and $N_{sim}$. From Eq. (4), it follows that the larger is the $N_{sim}$, the better is the $\bar{\varepsilon}$. Hence, the best average $\bar{\varepsilon} = 0.015946$ is obtained at $N_{sim} = 2^{19}$, and the $\bar{\varepsilon}$ obtained at $N_{sim} = 2^{19}$ is better than that obtained using smaller $N_{sim}$ for all $\delta$, for example, $\bar{\varepsilon}_{BAT-MCS} = 0.000346$ and $0.000120$ at $N_{sim} = 2^{17}$ and $2^{19}$, respectively, for $\delta = 1$.

**Table 9** and **Table 10** show that for a fixed $N_{sim}$, a smaller $\delta$ gives a larger average $n_{sim}(S)$ for all super vectors $S \in F_{\|S\|}$, thus obtaining better results. Hence, the best average $\bar{\varepsilon} = 0.001058$ for all $p$ is obtained at $\delta = 1$.

The dark cells in **Table 12** list the average values of $n_{sim}(S)$ from **Table 9** and **Table 10** that are less than 5, and these values may not have enough samples to describe the entire space solution completely. Hence, there is a gap between the values of dark cells and normal cells.

For each $N_{sim}$, there is at least one and up to three $\bar{\varepsilon}_{BAT-MCS}$ values that are better than $\bar{\varepsilon}_{MCS}$ at $\delta = 1$, 3, and 5. This further confirms the conclusion obtained from **Table 11** that the proposed algorithm gives better results than the MCS for lower $\delta$ than larger $\delta$.

Table 12. Results of $\bar{\varepsilon}$ for different $\delta$ and $N_{sim}$.

| $\delta$ | $2^{11}$ | $2^{13}$ | $2^{15}$ | $2^{17}$ | $2^{19}$ | avg |
|---|---|---|---|---|---|---|
| 0 | 0.004836 | 0.003728 | 0.000758 | 0.000663 | 0.000209 | 0.002039 |
| 1 | **0.001585** | 0.001988 | 0.001254 | **0.000346** | 0.000120 | **0.001058** |
| 3 | 0.003435 | 0.002475 | **0.000647** | 0.000553 | 0.000269 | 0.001476 |
| 5 | 0.002175 | **0.001779** | 0.001428 | 0.000396 | **0.000100** | 0.001176 |
| 7 | 0.009599 | 0.004409 | 0.001115 | 0.000737 | 0.000216 | 0.003215 |
| 9 | 0.033244 | 0.013252 | 0.001479 | 0.001086 | 0.000206 | 0.009854 |
| 11 | 0.355402 | 0.045336 | 0.007117 | 0.003070 | 0.000461 | 0.082277 |
| 13 | 1.482598 | 0.245925 | 0.085338 | 0.011128 | 0.001114 | 0.365220 |
| 15 | 1.837844 | 1.298944 | 0.365905 | 0.076730 | 0.006269 | 0.717139 |
| 17 | 1.997551 | 1.659655 | 1.235327 | 0.311806 | 0.056434 | 1.052155 |
| 19 | 0.703398 | 0.671208 | 0.624383 | 0.548766 | 0.110006 | 0.531552 |
| avg | 0.584697 | 0.358973 | 0.211341 | 0.086844 | **0.015946** | |

### 5.2.3 Average Standard Deviation $\bar{\sigma}$ of $R_{MCS}$ and $R_{BAT\text{-}MCS}$

**Table 13** and **Table 14** list the $\bar{\sigma}$ of the obtained $R_{MCS}$ and $R_{BAT\text{-}MCS}$ for different $\delta$ and $p$ and different $\delta$ and $N_{sim}$ using the MCS and proposed self-adaptive BAT-MCS, respectively.

Unlike in **Table 11** and **Table 12**, it is clear in both **Table 13** and **Table 14** that $\delta = 19$ always gives better $\bar{\sigma}$ than smaller $\delta$. This is a consequence of Eq. (27). It implies that the variance between the proposed algorithm and the MCS increases with an increase in $\delta$.

Moreover, the best average $\bar{\sigma}$ is 0.002934, obtained at $p = 0.9$ (**Table 13**), and 0.001405, obtained at $N_{sim} = 2^{19}$ (**Table 14**). The former may be a result of the characteristics shown in **Figure 4**, and the latter results from Eq. (4) and Eq. (25), that is, a larger $N_{sim}$ has a smaller $\bar{\sigma}$.

From the above discussions, it is clear that the proposed algorithm has a robust estimator because of its low variance.

Table 13. Standard deviations for different $\delta$ and $p$.

| $\delta$ | 0.1 | 0.3 | 0.5 | 0.7 | 0.9 | avg |
|---|---|---|---|---|---|---|
| 0 | 0.000596 | 0.003539 | 0.003554 | 0.001879 | 0.000347 | 0.001983 |
| 1 | 0.000612 | 0.003319 | 0.004111 | 0.002007 | 0.000361 | 0.002082 |
| 3 | 0.000543 | 0.002787 | 0.002916 | 0.001595 | 0.000286 | 0.001625 |
| 5 | 0.000566 | 0.002980 | 0.002956 | 0.001563 | 0.000229 | 0.001659 |
| 7 | 0.000485 | 0.002966 | 0.003117 | 0.001634 | 0.000260 | 0.001692 |
| 9 | 0.000488 | 0.002540 | 0.003159 | 0.001518 | 0.000289 | 0.001599 |
| 11 | 0.000447 | 0.002466 | 0.003097 | 0.001385 | 0.000275 | 0.001534 |
| 13 | 0.000407 | 0.001794 | 0.001404 | 0.001194 | 0.000250 | 0.001010 |
| 15 | 0.000392 | 0.001006 | 0.000668 | 0.000977 | 0.000229 | 0.000654 |
| 17 | 0.000279 | 0.000523 | 0.000316 | 0.000489 | 0.000222 | 0.000366 |
| 19 | **0.000226** | **0.000237** | **0.000083** | **0.000253** | **0.000186** | **0.000197** |
| avg | 0.005039 | 0.024157 | 0.025383 | 0.014493 | **0.002934** | |

Table 14. Standard deviations for different $\delta$ and $N_{sim}$.

| $\delta$ | $2^{11}$ | $2^{13}$ | $2^{15}$ | $2^{17}$ | $2^{19}$ | avg |
|---|---|---|---|---|---|---|
| 0 | 0.023924 | 0.013523 | 0.006838 | 0.003665 | 0.001624 | 0.009915 |
| 1 | 0.028306 | 0.012346 | 0.006542 | 0.003244 | 0.001613 | 0.010410 |
| 3 | 0.019739 | 0.011419 | 0.005203 | 0.002863 | 0.001412 | 0.008127 |
| 5 | 0.020445 | 0.011356 | 0.005272 | 0.002925 | 0.001478 | 0.008295 |
| 7 | 0.021986 | 0.010834 | 0.005591 | 0.002498 | 0.001401 | 0.008462 |
| 9 | 0.020792 | 0.009916 | 0.005446 | 0.002470 | 0.001342 | 0.007993 |
| 11 | 0.019218 | 0.009958 | 0.005171 | 0.002681 | 0.001325 | 0.007671 |
| 13 | 0.006552 | 0.009215 | 0.005373 | 0.002681 | 0.001421 | 0.005049 |
| 15 | 0.003724 | 0.004374 | 0.004180 | 0.002607 | 0.001475 | 0.003272 |
| 17 | 0.001410 | 0.002153 | 0.001750 | 0.002556 | 0.001272 | 0.001828 |
| 19 | **0.000693** | **0.001143** | **0.001165** | **0.000835** | **0.001089** | **0.000985** |
| avg | 0.015163 | 0.008749 | 0.004776 | 0.002639 | **0.001405** | |

## 5.2.4 Average Run Time $\bar{T}$

**Table 15** and **Table 16** list the average runtime $\bar{T}$ required to obtain the $R_{MCS}$ and $R_{BAT\text{-}MCS}$ for each run. The best and best average $\bar{T}$ are both obtained $p = 0.1$ and $\delta = 13$ results in the best average $\bar{T}$. In addition, the proposed algorithm is faster than the MCS for $\delta < 17$.

**Table 15** shows that $\delta = 13$ always has the best $\bar{T}$ and the best average $\bar{T}$ for different $p$. **Table 16** shows that the best and best average $\bar{T}$ for all $\delta$ are both obtained at $N_{sim} = 2^{11}$, which was the smallest $N_{sim}$ tested because a smaller $N_{sim}$ requires less runtime. However, the $\delta$ corresponding to the best $\bar{T}$ is different for various $N_{sim}$ values. For example, the best $\bar{T} = 0.002000$ occurs at $\delta = 7$ for $N_{sim} = 2^{11}$, but the best $\bar{T} = 0.006867$ occurs at $\delta = 11$ for $N_{sim} = 2^{13}$.

Thus, the proposed algorithm also outperforms the MCS in the runtime for $\delta < 17$ for all $p$ and all $N_{sim}$.

Table 15. $\bar{T}$ for different $\delta$ and $p$.

| $\delta$ | 0.1 | 0.3 | 0.5 | 0.7 | 0.9 | avg |
|---|---|---|---|---|---|---|
| 0 | 0.034900 | 0.055187 | 0.070947 | 0.060913 | 0.047633 | 0.053916 |
| 1 | 0.035867 | 0.035233 | 0.040853 | 0.041487 | 0.033173 | 0.037323 |
| 3 | 0.027727 | 0.033160 | 0.037087 | 0.033507 | 0.030200 | 0.032336 |
| 5 | 0.024553 | 0.028547 | 0.032193 | 0.030067 | 0.027427 | 0.028557 |
| 7 | 0.022627 | 0.026467 | 0.029320 | 0.026927 | 0.024547 | 0.025977 |
| 9 | 0.019833 | 0.022980 | 0.025707 | 0.024093 | 0.022040 | 0.022931 |
| 11 | 0.017393 | 0.020480 | 0.022467 | 0.021120 | 0.020007 | 0.020293 |
| 13 | **0.016540** | **0.019747** | **0.021467** | **0.020447** | **0.018153** | **0.019271** |
| 15 | 0.021253 | 0.023993 | 0.027613 | 0.025293 | 0.023013 | 0.024233 |
| 17 | 0.051340 | 0.053753 | 0.059427 | 0.054647 | 0.051727 | 0.054179 |
| 19 | 0.170953 | 0.173300 | 0.176967 | 0.175300 | 0.172780 | 0.173860 |
| avg | **0.040272** | 0.044804 | 0.049459 | 0.046709 | 0.042791 | |

Table 16. $\bar{T}$ for different $\delta$ and $N_{sim}$.

| $\delta$ | $2^{11}$ | $2^{13}$ | $2^{15}$ | $2^{17}$ | $2^{19}$ | avg |
|---|---|---|---|---|---|---|
| 0 | 0.002833 | 0.011600 | 0.063967 | 0.254467 | 1.015033 | 0.269580 |
| 1 | 0.002667 | 0.010400 | 0.044767 | 0.181633 | 0.693600 | 0.186613 |
| 3 | 0.002733 | 0.010467 | 0.037900 | 0.151933 | 0.605367 | 0.161680 |
| 5 | 0.002300 | 0.008533 | 0.034133 | 0.133967 | 0.535000 | 0.142787 |
| 7 | **0.002000** | 0.007733 | 0.030933 | 0.122467 | 0.486300 | 0.129887 |
| 9 | 0.006667 | 0.011500 | **0.027167** | 0.105967 | 0.421967 | 0.114653 |
| 11 | 0.002500 | **0.006867** | 0.029933 | 0.098467 | 0.369567 | 0.101467 |
| 13 | 0.012233 | 0.016467 | 0.032533 | **0.092667** | 0.327867 | **0.096353** |
| 15 | 0.047200 | 0.050167 | 0.068433 | 0.121233 | **0.318800** | 0.121167 |
| 17 | 0.212833 | 0.214800 | 0.223800 | 0.271433 | 0.431600 | 0.270893 |
| 19 | 0.792700 | 0.794067 | 0.863100 | 0.886267 | 1.010367 | 0.869300 |
| avg | **0.098788** | 0.103873 | 0.132424 | 0.220045 | 0.565042 | |

The discussed experimental results indicate that the proposed self-adaptive BAT-MCS outperforms the MCS in terms of quality, variance, and runtime for suitable $\delta$ values.

## 6. CONCLUSIONS

This paper presented a novel MCS, called the self-adaptive BAT-MCS, to calculate binary-state network reliability. Experimental results showed that the proposed self-adaptive BAT-MCS based on BAT, MCS, proposed lower/upper bounds for super vectors, and proposed self-adaptive simulation-number algorithm outperformed the MCS in terms of quality, variance, and runtime for suitable $\delta$ values.

Compared with the traditional MCS, the proposed self-adaptive BAT-MCS is simple, efficient, and robust in terms of time complexity, solution quality, and variance as shown experimentally. Therefore, from computational, pragmatic, and theoretical perspectives, the proposed BAT-MCS is superior to the BAT, MCS, and reliability-bound algorithms for calculating the reliability of binary-state networks. The future work is to extend the proposed algorithm to the multi-state flow networks reliability problems [44-54].

## ACKNOWLEDGMENT

This research was supported in part by the Ministry of Science and Technology, R.O.C. under grant MOST 110-2221-E-007-107-MY3. This article was once submitted to arXiv as a temporary submission that was just for reference and did not provide the copyright.

**REFERENCES**

[1] https://en.wikipedia.org/wiki/Monte_Carlo_method

[2] S. Y. Chen, K. C. Hsu, and C. M. Fan (2021). "Improvement of generalized finite difference method for stochastic subsurface flow modeling", Journal of Computational Physics, 429, 110002.

[3] X. Jia, P. Ziegenhein, and S. B. Jiang (2014). "GPU-based high-performance computing for radiation therapy", Physics in Medicine & Biology, 59(4), R151.

[4] Z. Zhongming, L. Linong, Z. Wangqiang, and L. Wei (2013). "Climate change 2013: The physical science basis".

[5] Ojeda P., Garcia M. E., A. Londoño, and N. Y. Chen (2009). "Monte Carlo simulations of proteins in cages: influence of confinement on the stability of intermediate states", Biophysical journal, 96(3), 1076-1082.

[6] A. J. Cassey, and B. O. Smith (2014). "Simulating confidence for the Ellison–Glaeser index", Journal of Urban Economics, 81, 85-103.

[7] T. Jakl (2011). "Arimaa challenge–comparission study of MCTS versus alpha-beta methods".

[8] L. Szirmay-Kalos (2008). "Monte Carlo Methods in Global Illumination-Photo-realistic Rendering with Randomization", VDM Verlag.

[9] L. D. Stone, T. M. Kratzke, and J. R. Frost (2011). "Search modeling and optimization in USCG's search and rescue optimal planning system (SAROPS)", In Proceedings of the Safer Seas Conference, Brest, France, Vol. 10.

[10] D. J. Arenas, L. A. Lett, H. Klusaritz, and A. M. Teitelman (2017). "A Monte Carlo simulation approach for estimating the health and economic impact of interventions provided at a student-run clinic", PloS one, 12(12), e0189718.

[11] D. Kakadia, and J. E. Ramirez-Marquez (2020). "Quantitative approaches for optimization of user experience based on network resilience for wireless service provider networks", Reliability Engineering & System Safety, 193, 106606.

[12] C. Lin, L. Cui, D. W. Coit, and M. Lv (2017). "Performance Analysis for a Wireless Sensor Network of Star Topology with Random Nodes Deployment", Wireless Personal Communications, 97(3), 3993-4013.

[13] T. Aven (1987). "Availability evaluation of oil/gas production and transportation systems",

Reliability engineering ,18(1), 35-44.

[14] W. C. Yeh, and J. S. Lin (2018). "New parallel swarm algorithm for smart sensor systems redundancy allocation problems in the Internet of Things", The Journal of Supercomputing, 74(9), 4358-4384.

[15] L.A. Jason, and E. Stevens (2017). "The Reliability and Reciprocity of a Social Network Measure", Alcoholism Treatment Quarterly, 35(4), 317-327.

[16] https://www.statista.com/statistics/268136/top-15-countries-based-on-number-of-facebook-users/

[17] C. J. Colbourn (1987). "The combinatorics of network reliability", Oxford University Press, Inc.

[18] D. R. Shier (1991). "Network reliability and algebraic structures", Clarendon Press.

[19] Y. F. Niu, and F. M. Shao (2011). "A practical bounding algorithm for computing two-terminal reliability based on decomposition technique", Computers & Mathematics with Applications, 61(8), 2241-2246.

[20] D. W. Coit, and E. Zio (2018). "The evolution of system reliability optimization", Reliability Engineering & System Safety, 192, 106259.

[21] W. C. Yeh, Y. C. Lin, and Y. Y. Chung (2010). "Performance Analysis of Cellular Automata Monte Carlo Simulation for estimating Network Reliability", Expert Systems with Applications, 37(5), 3537-3544.

[22] W. C. Yeh, L. Cao, and J. S. Jin (2012). "A Cellular Automata Hybrid Quasi-random Monte Carlo Simulation for Estimating the One-to-all Reliability of Acyclic Multi-state Information Networks", International Journal of Innovative Computing, Information and Control, 8(3), 2001-2014.

[23] J. E. Ramirez-Marquez (2015). "Assessment of the transition-rates importance of Markovian systems at steady state using the unscented transformation", Reliability Engineering & System Safety, 142, 212-220.

[24] W. C. Yeh and C. H. Lin, "A Squeeze Response Surface Methodology for Finding Symbolic Network Reliability Functions", IEEE Transactions on Reliability, 58(2), 374-382.

[25] W. C. Yeh (2016). "A Squeezed Artificial Neural Network for the Symbolic Network Reliability Functions of Binary-State Networks", IEEE Transactions on Neural Networks and Learning Systems, 28 (11), 2822-2825.

[26] J. M. Hammersley, and D. C. Handscomb (1964). "Monte Carlo Methods", Chapman and Hall, London.

[27] Magalhães F., Monteiro J., J. A. Acebrón, and J. R. Herrero (2022). "A distributed Monte Carlo based linear algebra solver applied to the analysis of large complex networks", Future Generation Computer Systems, 127, 320-330.

[28] W. C. Yeh (2021). "Novel binary-addition tree algorithm (BAT) for binary-state network reliability problem", Reliability Engineering & System Safety, 208, 107448.

[29] W. C. Yeh, Z. Hao, M. Forghani-Elahabad, G. G. Wang, and Y. L. Lin (2021). "Novel Binary-Addition Tree Algorithm for Reliability Evaluation of Acyclic Multistate Information Networks", Reliability Engineering & System Safety, 210, 107427.

[30] W. C. Yeh (2021). "A Quick BAT for Evaluating the Reliability of Binary-State Networks", Reliability Engineering & System Safety, 216, 107917.

[31] Y. Z. Su, and W. C. Yeh (2020). "Binary-Addition Tree Algorithm-Based Resilience Assessment for Binary-State Network Problems", Electronics, 9(8), 1207.

[32] W. C. Yeh, and C. C. Kuo (2020). "Predicting and modeling wildfire propagation areas with BAT and maximum-state PageRank", Applied Sciences, 10(23), 8349.

[33] W. C. Yeh, E. Lin, and C. L. Huang (2021). "Predicting Spread Probability of Learning-Effect Computer Virus", Complexity, 2021, 6672630.

[34] Z. Hao, W. C. Yeh, J. Wang, G. G. Wang, and B. Sun (2019). "A quick inclusion-exclusion technique", Information Sciences, 486, 20-30.

[35] M. J. Zuo, Z. Tian, and H. Z. Huang (2007). "An efficient method for reliability evaluation of multistate networks given all minimal path vectors", IIE transactions, 39(8), 811-817.

[36] G. Levitin (2005). "The universal generating function in reliability analysis and optimization", Springer.

[37] W. C. Yeh (1998). "A revised layered-network algorithm to search for all d-minpaths of a limited-flow acyclic network", IEEE Transactions on Reliability, 47(4), 436-442.

[38] Y. Niu, Z. Gao, and H. Sun (2017). "An improved algorithm for solving all *d*-MPs in multi-state networks", Journal of Systems Science and Systems Engineering, 26(6), 711-731.

[39] W. C. Yeh (2021). "Novel Algorithm for Computing All-Pairs Homogeneity-Arc Binary-State Undirected Network Reliability", Reliability Engineering & System Safety, 216, 107950.

[40] Z. Hao, W. C. Yeh, S. Y. Tan (2021). "One-batch Preempt Deterioration-effect Multi-state Multi-rework Network Reliability Problem and Algorithms", Reliability Engineering & System Safety, 215, 107883.

[41] W. C. Yeh (2021). "New Binary-Addition Tree Algorithm for the All-Multiterminal Binary-State Network Reliability Problem", arXiv preprint arXiv:2111.10818.

[42] A. Satyanarayana, and Mark K. Chang (1983). "Network Reliability and the Factoring Theorem", Networks, 13(1), 107-120.

[43] N. Metropolis, and S. Ulam (2012). "The Monte Carlo Method", Journal of the American Statistical Association, 44(247), 335-342.

[44] P.C. Chang (2017). "A simulation analysis of the impact of finite buffer storage on manufacturing system reliability", Simulation Modelling Practice and Theory 70, 149-158.

[45] https://sites.google.com/view/wcyeh/source-code-download.

[46] C. F. Huang (2020). "System reliability for a multi-state distribution network with multiple terminals under stocks", Annals of Operations Research: 1-14.

[47] https://weibull.com/hotwire/issue3/hottopics3.htm

[48] J. Faulin, A. A. Juan, S. Martorell, and J.-E. Ramírez-Márquez (2010). Simulation Methods for Reliability and Availability of Complex Systems, DOI: 10.1007/978-1-84882-213-9, Springer Series in Reliability Engineering book series.

[49] J.-E. Ramírez-Márquez and D. W. Coit (2010). "A Monte-Carlo simulation approach for approximating multi-state two-terminal reliability", Reliability Engineering & System Safety, 87(2), 253-264.

[50] P. Zhu, Y. Guo, J. Han, and F. Lombardi (2017). "On the Approximate Reliability of Multi-state Two-Terminal Networks by Stochastic Analysis", IET Networks, 6(5), DOI: 10.1049/iet-net.2017.0033.

[51] J.-E. Ramírez-Márquez, D. W. Coit, and M. Tortorella (2006). "A generalized multistate-based path vector approach to multistate two-terminal reliability", IIE Transactions 38(6), 477-488.

[52] E. Zio and N. Pedroni (2010). "Reliability Estimation by Advanced Monte Carlo Simulation. Editors: Faulin, Juan, Martorell, Ramirez-Marquez. Simulation Methods for Reliability and Availability of Complex Systems, 3-39, Springer Series in Reliability Engineering.

[53] J.-E. Ramírez-Márquez and D. W. Coit (2005). "Composite importance measures for multi-state systems with multi-state components", IEEE Transactions on Reliability, 54(3), 517-529.

[54] B. M. Ayyub and R. H. Mccuen (1995) Simulation-Based Reliability Methods. In: Sundararajan C. (eds) Probabilistic Structural Mechanics Handbook. Springer, Boston, MA. https://doi.org/10.1007/978-1-4615-1771-9_4.